\documentclass[%
 notitlepage,
 reprint,
 superscriptaddress,
 amsmath,amssymb,
 aps,
]{revtex4-1}

\usepackage{lipsum}
\usepackage{graphicx}
\usepackage{dcolumn}
\usepackage{bm}

\usepackage{epstopdf}

\newcommand{\bra}[1]{\langle #1 \vert}
\newcommand{\ket}[1]{\vert #1 \rangle}

\usepackage{hyperref}

\begin{document}

\preprint{APS/123-QED}

\title{Disorder-safe entanglement transfer through ladder qubit chains}

\author{Guilherme M. A. Almeida}
\email{gmaalmeida.phys@gmail.com}
\affiliation{%
 Instituto de F\'{i}sica, Universidade Federal de Alagoas, 57072-900 Macei\'{o}, AL, Brazil
}%
\author{Andre M. C. Souza}
\affiliation{%
 Departamento de F\'{i}sica, Universidade Federal de Sergipe, 49100-000 S\~{a}o Crist\'{o}v\~{a}o, SE, Brazil
}%
\author{Francisco A. B. F. de Moura}
\affiliation{%
 Instituto de F\'{i}sica, Universidade Federal de Alagoas, 57072-900 Macei\'{o}, AL, Brazil
}%
\author{Marcelo L. Lyra}
\affiliation{%
 Instituto de F\'{i}sica, Universidade Federal de Alagoas, 57072-900 Macei\'{o}, AL, Brazil
}%

\date{\today}

\begin{abstract}
We study an entanglement transfer protocol in a two-leg ladder spin-1/2 chain in the presence of disorder.
In the regime where on-site energies and the intrachain couplings follow aproximately constant proportions
locally, we set up a scheme for high-fidelity state transfer via a disorder-protected subspace 
wherein fluctuations in the parameters do not depend on
the global disorder of the system, accounted by $W$. 
Moreover, we find that the leakage of information from that subspace is actually suppressed upon increasing $W$ and thus
the transfer fidelity, evaluated through the entanglement concurrence at the other end of the chain,
builds up
with the disorder strength. 
\end{abstract}

\maketitle


\section{\label{sec1}Introduction}

In the past decade, 1D spin chains have been regarded as potential quantum communication platforms for a wide variety of tasks (see \cite{apollarorev,nikobook} and references within).
In standard quantum-state transfer protocols \cite{bose03}, for instance, the chain must be manufactured in such a way
an arbitrary qubit can be faithfully sent from one point to another at some (preferably small) prescribed time following the natural underlying Hamiltonian dynamics of the system.

To do so, a handful of schemes have have been put forward since the original proposal in Ref. \cite{bose03}, some relying on fully-engineered couplings \cite{christandl04,plenio04,niko04epl} -- thereby yielding perfect transfer through arbitrary distances --, dual-rail encoding \cite{burgarth05}, strong local magnetic fields \cite{lorenzo13,lorenzo15}, and weak-end couplings \cite{wojcik05,*wojcik07,li05, almeida16,almeida18pra} to name a few.       

Given the possibility of experimental errors in the manufacturing process of the chain and 
that one is willing not to interfere with the channel while it is operating in order to avoid
decoherence and losses, disorder stands out as a major threat to the performance of the protocol.
This fact alone has motivated several studies on the
influence of static fluctuations in the parameters of the chain 
over the state transfer fidelity \cite{allcock09,dechiara05,fitzsimons05,burgarth05,tsomokos07, petrosyan10,yao11,zwick11,*zwick12,bruderer12,kay16,almeida17-1,almeida18pla,almeida18ann}.  

It is pretty well established that 1D and 2D single-particle hopping models
display Anderson localization for any degree of uncorrelated disorder \cite{anderson58,*abrahams79}.
A very rich cross-over between localization and delocalization, though, can be found in 
chains displaying certain kinds of correlated disorder \cite{dunlap90,*phillips91,demoura98,izrailev99,kuhl00,izrailev12rev}. For instance, 
it was shown in \cite{demoura98} that long-range correlated disorder induces the appearance
of a band of extended states with sharp mobility edges thereby indicating a metal-insulator transition.
Very recently, we have explored this breakdown of Anderson localization in order to 
carry out quantum-state transfer protocols \cite{almeida17-1,almeida18pla,almeida18ann}.
Another kind of configuration that deserves attention is quasi-1D models such as ladder networks. 
In \cite{sil08}, it was reported that a two-leg Aubry-Andr\'{e} model
displays a metal-insulator transition with multiple 
mobility edges. They also 
put forward the possibility of spanning a band of 
disorder-free states coexisting with exponentially-localized modes given the
on-site energies and interchain hopping strenghts follow constant proportions along the ladder \cite{sil08prb}. de Moura et al. further found out a novel level-
spacing statistics associated to it \cite{demoura10}. A generalized version of 
this wavefunction delocalization engineering for  
$N$-leg ladder systems has also been put forward \cite{rodriguez12}.
The inteplay between channels featuring different degrees and/or types of disorder
has also been investigated \cite{zhang10,guo11}.
 
In this work we bring about the idea of disorder-free subspaces spanning
over a strongly disordered media \cite{sil08prb,rodriguez12}
into the context of quantum communication protocols. 
In particular, we aim to transmit entanglement with high fidelity from one end
of a two-leg ladder chain to the other in the presence of disorder. We outline the parameter conditions for which a disorder-free channel arises and how to properly encode the initial entangled state in order to use it. We further consider imperfections in this channel which can 
promote the 
leakage of information out of it. Our main result is that this effect can be avoided when 
we \textit{increase} the amount of disorder originally present in the system. This remarkable behavior paves the way for disorder-safe quantum communication protocols in engineered qubit chains.     
 
In the following, Sec. \ref{sec2} we introduce the Hamiltonian model and in Sec. \ref{sec3} 
we discuss the conditions for setting up a disorder-free channel. In Sec. \ref{sec4} we display our results for the entanglement transfer performance against disorder as well as we investigate the leakage of information out of the protected subspace. Our conclusions are drawn in Sec. \ref{sec5}.

\section{\label{sec2}Model and formalism}

Here, we deal with a two-leg ladder spin (qubit) chain of the $XX$ type, with $N$ sites each, described 
in terms of a free-fermion
Hamiltonian of the form $H = H^{(1)}+H^{(2)}+H_{I}$, with ($\hbar = 1$)
\begin{align} 
    H^{(j)} &= \sum_{n=1}^{N} \epsilon_{n,j}c_{n,j}^{\dagger}c_{n,j}
     + \sum_{n=1}^{N-1}J_{n,j}(c_{n+1,j}^{\dagger}c_{n,j}+\mathrm{H.c.}),\label{mainH1}\\
    H_{I} &= \sum_{n=1}^{N}\gamma_{n}(c_{n,2}^{\dagger}c_{n,1}+\mathrm{H.c.}),\label{mainH2}
\end{align}
where $c_{n,j}^{\dagger}$ ($c_{n,j}$) creates (removes) a fermion
at the $n$-th
site of the $j$-th chain ($j=1,2$), $\epsilon_{n,j}$ stands for its local potential, 
$J_{n,j}$ is the intra-chain hopping rate, and $\gamma_{n}$ is the inter-chain hopping rate. 
Throughout this paper we consider $J_{n,1} = J_{n,2}\equiv J_{n}$ and set $J = \mathrm{max} \lbrace J_{n} \rbrace \equiv 1$ as the energy unit.
Note that $H$ conserves the total number of excitations. 
Here we are interested in the single-excitation manifold spanned
by 
\begin{equation}\label{statenotation}
\ket{\textbf{n}}^{(j)}\equiv c_{n,j}^{\dagger}\bigotimes_{i=1}^{2}\ket{00\cdots 0}^{(i)},
\end{equation}
thereby forming a 2N-dimensional Hilbert space.  

We now allow disorder to occur 
on the on-site potentials $\epsilon_{n,j}$ and inter-chain coupling rates $\gamma_{n}$.
In particular, we assume these quantities 
to fall within a 
uniform random distribution in the 
interval $[- W,W]$, $W$ being 
the intensity of disorder.
Further, we consider $J_{n,1} = J_{n,2} \equiv J_{n}$. 

In \cite{sil08prb} (see also \cite{demoura10}) it was shown that 
when $\epsilon_{n,1}$, $\epsilon_{n,2}$, and $\gamma_{n}$ 
obey constant proportions between each other across the chain, 
one is able to choose an appropriate basis set that decouples both legs. Moreover, 
it is possible to turn one of them completely free of disorder \cite{sil08prb}. To see this happening, 
let us define 
\begin{equation} \label{newbasis}
    \ket{\textbf{n},\pm} = \dfrac{\ket{\textbf{n}}^{(1)}\pm \ket{\textbf{n}}^{(2)}}{\sqrt{2}},
\end{equation}
and rewrite Hamiltonian $H$ in terms of these states, to get
\begin{align} \label{effH}
    H &=\sum_{\mu = \pm} \left[ \sum_{n=1}^{N} \widetilde{\epsilon}_{n,\mu}\ket{\textbf{n},\mu}
    \bra{\textbf{n},\mu}  +  \sum_{n=1}^{N-1}J_{n}  
    (\ket{\textbf{n}+\textbf{1},\mu}\bra{\textbf{n},\mu}+\mathrm{H.c.}) \right] \nonumber \\
    &\,\,\,\,\,\, + \sum_{n=1}^{N}\widetilde{\gamma}_{n}(\ket{\textbf{n},+}
    \bra{\textbf{n},-}+\mathrm{H.c.}),
\end{align}
with
\begin{align}
    \widetilde{\epsilon}_{n,\pm} &= \dfrac{\epsilon_{n,1}
    +\epsilon_{n,2}}{2} \pm \gamma_{n}, \label{eff_ep} \\
    \widetilde{\gamma}_{n} &= \dfrac{\epsilon_{n,1}
    -\epsilon_{n,2}}{2}, \label{eff_gamma}
\end{align}
being the potentials and inter-chain coupling rates, respectively, for the effective ladder with both legs extending over $\ket{\textbf{n},+}$ and $\ket{\textbf{n},-}$.

\section{\label{sec3}Disorder-free subspace}

From Hamiltonian (\ref{effH}), we readily note that  
setting $\epsilon_{n,1} = \epsilon_{n,2}$ 
leads to the decoupling of both legs.
In addition, when $\epsilon_{n,1} = \gamma_{n}$
then the anti-symmetric branch 
takes $\widetilde{\epsilon}_{n,-} =0$. 
The availability of a ordered subspace embedded in a strongly
disordered media is very appealing when it comes to, e.g., 
running pre-engineered 
quantum-state transfer protocols \cite{bose03}.
Suppose we have an imperfect (single-leg) chain 
due to uncorrelated on-site fluctuations. By 
generating another copy of it and linking them 
up one may find a clean, disorder-free
quantum communication channel by properly enconding
the qubit [e.g. following Eq. (\ref{newbasis})] to be sent through, no matter how strong $W$ is.

A possible issue that might set in, though, is that the second (backup) leg may not be a legit copy
of the first one. That would 
keep disorder in the channel
as well as promote the leakage of information
from $\ket{\textbf{n},-}$ into $\ket{\textbf{n},+}$.
Still, 
if we allow for small deviations in, say, $\epsilon_{n,2}$
around $\epsilon_{n,1}$, it is possible to keep the channel reasonably safe. 
For instance, let 
$\epsilon_{n,2} = \epsilon_{n,1} + \delta_{n}$,
with $\delta_{n}$ 
being another random number, uniformly distributed within
$[-\Delta,\Delta]$ such that $\Delta \ll W$. 
%
By looking at Eqs. (\ref{eff_ep}) and (\ref{eff_gamma}), 
we now have $\widetilde{\gamma}_{n} = -\delta_{n} / 2$ and
$\widetilde{\epsilon}_{n,-} =\delta_{n} / 2$. 
Therefore, disorder in leg $\lbrace \ket{\textbf{n},-}\rbrace$ is solely
weighted by $\Delta$ and not by $W$, 
so that the latter can be arbitrarily large.
If $\Delta\neq 0$ there will be leakage into $\lbrace \ket{\textbf{n},+}\rbrace$, this subspace now acting as an disordered ``environment'' with on-site energies given by 
$\widetilde{\epsilon}_{n,+} =(4\epsilon_{n,1}+ \delta_{n}) / 2$.
Note that we are still considering $\gamma_{n} = \epsilon_{n,1}$. Small deviations from it would
not affect $\widetilde{\gamma}_{n}$, but the on-site potentials $\widetilde{\epsilon}_{n,+}$ and $\widetilde{\epsilon}_{n,-}$ with a small shift. Because of that, here disorder will be ultimately set by $W$ (global disorder) and $\Delta$ in the regime $\Delta \leq W$ with $\Delta \ll J$.

%

\section{\label{sec4}Results and discussion}

\subsection{Entanglement transfer}
 
Now, suppose 
Alice has access to the first ``cell'' of the ladder and is willing to 
send some amount of entanglement to Bob at 
at the other end of the ladder
relying only upon the natural Hamiltonian dynamics of the system \cite{bose03}.
In order to make use of the disorder-protected subspace as discussed in the previous section, 
a bipartite entangled state of the Bell type can be properly prepared in the following way. A single-excitation (spin up)
is initially set by Alice in one of her sites of domain $\ket{01}_{A}$. By further 
applying the Hadamard gate $H_{\mathrm{gate}}$ at her second qubit followed by a CNOT gate controlled by 
the same one (being $\ket{0}$) she gets
\begin{align}
\ket{\phi}_{A} &= CNOT ( I \otimes H_{\mathrm{gate}})\ket{01}_{A}  \nonumber \\
&= CNOT \ket{0}_{A}\left(\frac{\ket{0}_{A}-\ket{1}_{A}}{\sqrt{2}} \right) \nonumber  \\
&= \frac{\ket{10}_{A}-\ket{01}_{A}}{\sqrt{2}},
\end{align}
that is, a maximally entangled Bell state. The \textit{whole} system is then initialized in
$\ket{\psi(0)} = \ket{\phi}_{A}\ket{0\cdots 0}^{(1)}\ket{0\cdots 0}^{(2)} = \ket{\mathbf{1},-}$ [cf. Eqs. (\ref{statenotation}) and \ref{newbasis}], which
can be thought as a particular case of the dual-rail encoding scheme \cite{burgarth05}.

We are now to quantify the amount of entangled to reach Bob's cell through unitary evolution of Hamiltonian (\ref{effH}), $U(t) \equiv e^{-iHt}$. 
For this, we resort to the so-called concurrence \cite{wootters98} which accounts for the entanglement shared between two qubits
in any arbitrary mixed state. For single-particle states, all the input we need is the wave function amplitude
of both qubits of interest, namely $C(t) = 2|f_{N}^{(1)}(t)f_{N}^{(2)}(t)|$, where
$f_{N}^{(j)} \equiv \langle \mathbf{N}^{(j)}  \vert \psi(t) \rangle$ is the transition amplitude 
to the last site of the $j$-th leg. 
For a separable (fully-entangled) state, this quantity reads $C=0$ ($C=1$).
Note that the transfer performance will be ruled by the likelihood of having $\ket{\psi(\tau)} \approx \ket{\mathbf{N},-}$ at a given
time $\tau$. We thus need to come up with some coupling scheme 
for carrying out high fidelity excitation transfer from one end of the chain to the other. 
Here, in particular, we choose the 
class of fully-engineered couplings used in perfect state transfer protocols \cite{christandl04},
$J_{n} = \sqrt{n(N-n)}$, with $n = 1,2,\ldots, N-1$. This scheme brings about a linear dispersion relation
thereby allowing for transmission
of quantum states with maximum fidelity (in an ordered system) in chains 
with arbitrary size at time $\tau = \pi N / (4J)$ \cite{christandl04}. Experimental realizations
of this configuration have been put forward in \cite{bellec12, chapman16}.

\begin{figure}[t!] 
\includegraphics[width=0.45\textwidth]{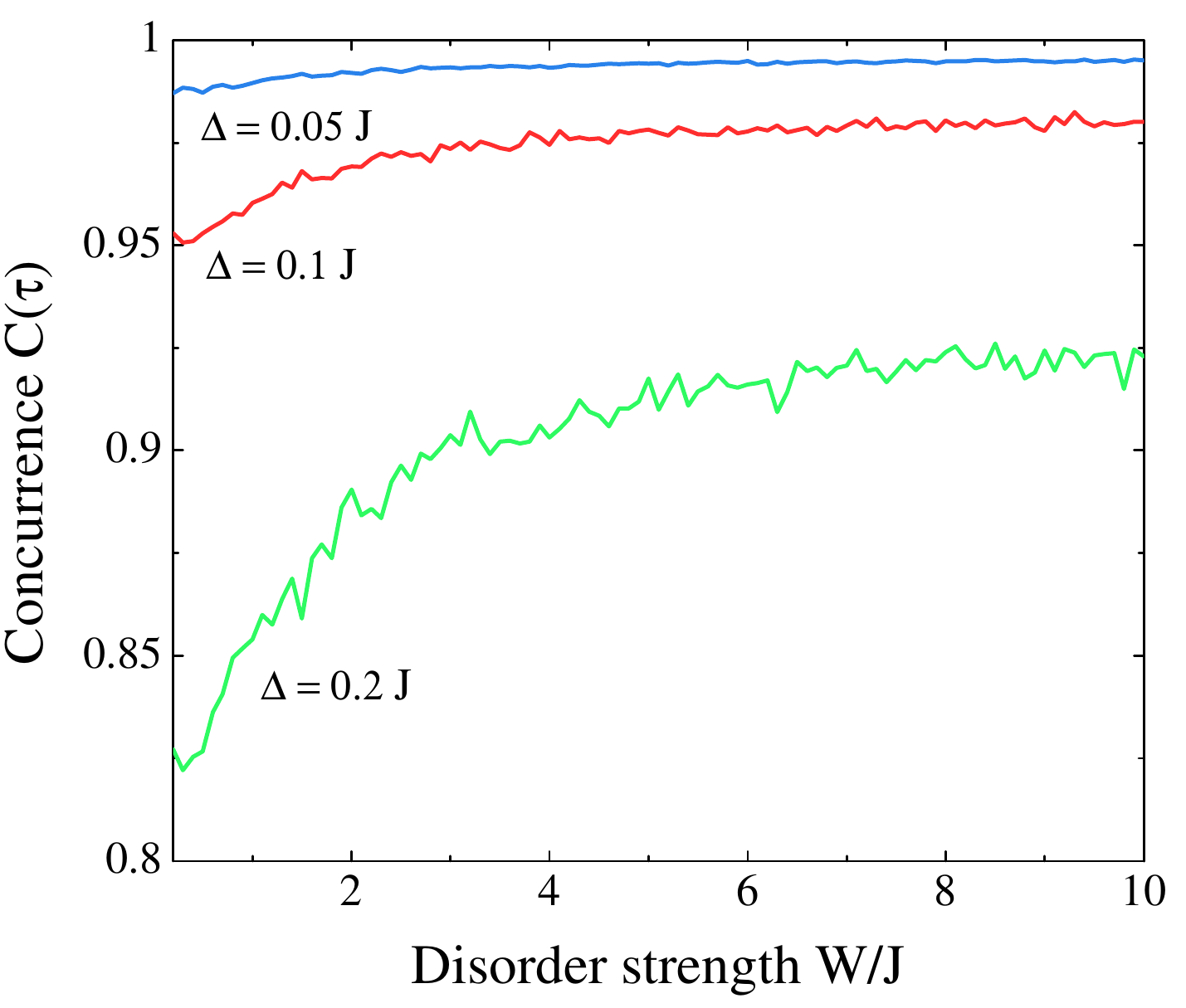}
\caption{\label{fig1} Two-qubit concurrence for the last cell of the ladder system,
$C(\tau)$, evaluated at time $\tau = \pi N /(4J)$ against (global) 
disorder strength $W/J$ (from $0.2$ to $10$) averaged over $100$ distinct realizations of it.
The coupling scheme was that of perfect quantum state transfer protocols \cite{christandl04}, 
$J_{n} = \sqrt{n(N-n)}$.
Plots were obtained by exact numerical diagonalization of Hamiltonian (\ref{effH})
for $N=30$ and $\Delta/J = 0.05$, $0.1$, and $0.2$.
The initial state was $\ket{\psi (0)} = \ket{\mathbf{1},-}$.
}
\end{figure}

Figure \ref{fig1} shows the disorder-averaged entanglement concurrence $C(t)$ versus disorder strength $W$ evaluated 
at $t = \tau$ for the encoded initial state $\ket{\psi (0)} = \ket{\mathbf{1},-}$. 
There we readily spot a very interesting behavior, namely that the entanglement transfer performance
actually gets better upon increasing $W$. 
At this point it is convenient to recall that
if $\Delta = 0$ then the dynamics takes place on a disorder-free subspace, namely an
effective ordered 1D chain (see beginning of Sec. \ref{sec3}). In that case, 
the concurrence would be maximum, $C(\tau) = 1$, entailing a perfect state transfer.
In Fig. 1 we see that a local detuning in each cell, $\Delta \neq 0$, 
lowers the transfer performance. This happens because an effective internal disorder has been induced
in branch $\lbrace \ket{\textbf{n},-}\rbrace$ at the same time information is leaking from it into $\lbrace \ket{\textbf{n},+}\rbrace$.
We also mention that these fluctuations affect the transfer time $\tau$.
On the other hand, upon increasing $W$, the concurrence is substantially recovered until saturating. 
At this regime, the global disorder $W$ no longer has influence on $C(\tau)$, rather, its 
saturated (averaged) value is set upon $\Delta$. 
The reason behind it is that the leakage 
is ultimately suppressed given $W$ is high enough as we 
are to discuss next.

 \subsection{Leakage dynamics}

\begin{figure}[t!] 
\includegraphics[width=0.45\textwidth]{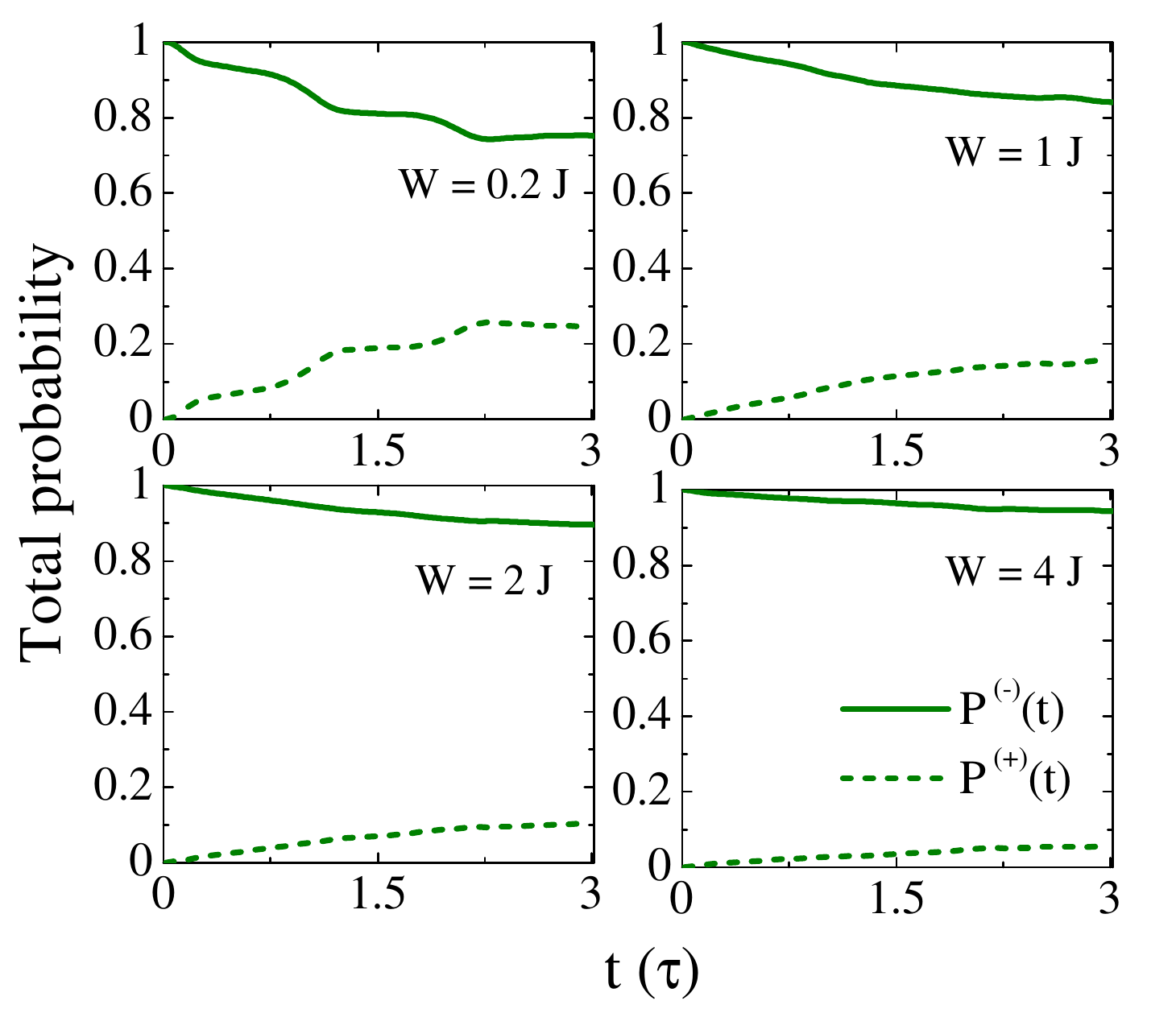}
\caption{\label{fig2} Time evolution of the total occupation probability in the negative branch, 
$P^{(-)}(t) \equiv \sum_{n}|\langle \textbf{n},- | \psi(t) \rangle|^2$ (solid lines), and in the positive one,
$P^{(+)}(t) = 1-P^{(-)}(t)$ (dashed lines),
for various disorder strengths $W$, averaged over $100$ realizations of disorder.
Time is expressed in units of $\tau = \pi N /(4J)$. 
Plots were obtained by exact numerical diagonalization of Hamiltonian (\ref{effH})
for $N=30$, $\Delta/J = 0.2$, and $J_{n} = \sqrt{n(N-n)}$.
The initial state was $\ket{\psi (0)} = \ket{\mathbf{1},-}$. 
}
\end{figure}

At this point, we are led to investigate the leakage of information
from the anti-symmetric branch and 
the influence of disorder strength $W$ in order to explain the transfer (concurrence) outcomes seen in Fig. \ref{fig1}. 
%
%

Before doing so, we shall get some intuition over the dynamics of disordered ladders
by looking at its physical (original) form [Eqs. (\ref{mainH1}) and (\ref{mainH2})]. 
For a moment, suppose 
the local energy detuning $\delta_{n} = \delta$ 
and $\gamma_{n} = \gamma$ for all $n$. 
If we set an initial state as 
any linear combination of, say, $\lbrace \ket{\mathbf{n}^{(1)}}\rbrace$, the overall occupation probability to remain 
in the first leg reads
$P^{(1)}(t) \equiv \cos^{2}(\gamma t)$
for $\delta = 0$ (see Appendix for details).
Thereby, 
the excitation
keeps oscillating back and forth between both legs
periodically whereas it goes by following its own intrachain dynamics. 
For $\delta \neq 0$, $P^{(1)}(t)$ still undergoes periodic oscillations but with smaller amplitude and faster rate (see Eq. (\ref{overallP}) of Appendix). The simple picture above tells us in advance that large detunings prevent
leakage of information from one leg to the other, as we would have expected intuitively.     

Let us now get back to the \textit{effective} ladder chain described by Hamiltonian (\ref{effH}), wherein the local cell detunings and interchain couplings 
follow disordered sequences along the array. 
In Figure \ref{fig2} we show the time evolution of 
$P^{(-)}(t) \equiv \sum_{n}|\langle \textbf{n},- | \psi(t) \rangle|^2$
for the same ladder configuration as in Fig. \ref{fig1} and initial state $\ket{\psi (0)} = \ket{\mathbf{1},-}$, averaged over many distinct realizations of disorder.
First and foremost, there it is clear 
that the disorder strength $W$ indeed prevents
the excitation to leak from subspace $\lbrace \ket{\textbf{n},-}\rbrace$ into $\lbrace \ket{\textbf{n},+}\rbrace$.
We shall also mention that the curves look smooth due to 
the disorder-averaging procedure. Each realization now displays a
non-periodic oscillatory behavior. 
For longer times, the (averaged) total probability reaches about a stationary value which
depends on both $\Delta$ and $W$. 

When $\Delta \sim W$ and the system is initialized in the effective leg 
featuring a very low amount of disorder (that is the negative branch), 
the excitation spreads out far away from the initial site and so it 
is very likely that it
will eventually find some 
local resonance -- a given cell with low detuning $\widetilde{\epsilon}_{n,+} \approx \widetilde{\epsilon}_{n,-}$ -- and therefore the excitation is capable of
making through the other leg. Chances are extremely low for this to happen upon increasing $W$ and thus
the excitation becomes trapped in the original leg for high enough $W \gg \Delta$. 

\section{\label{sec5}Conclusions}

We studied a entanglement transfer protocol set over a disordered two-leg ladder qubit chain.
By generating a disorder-free subspace, we carried out the protocol in the presence of local imperfections that lead to the leakage of information into the strongly-disordered manifold. We showed that increasing the disorder strength $W$ prevents such leakage thereby improving the concurrence (transfer) outcomes at the target location to a great extent. We further explained it by studying the leakage dynamics in detail. 

This rather surprising behavior shows us that disorder 
may be an essential ingredient to prevent dissipation.
Indeed, there has been considerable interest in
studying open system dynamics involving 
structured (such as disordered) environments 
\cite{lorenzo17,cosco18}.
In \cite{lorenzo17}, for instance -- by looking at the dynamics
of a single emitter coupled to an array of cavities
acting as the environment --
they reported that disorder pushes information back
to the emitter. They further characterized this information
backflow using proper non-Markovianity measures. 
Further extensions of our work may be taken along this direction. 

Another possibility is to setting up quantum communication protocols in $N$-leg ladders for which
there also exists some methods to induce a disorder-free subspace embedded within
a strongly-disordered scenario \cite{rodriguez12}.

 

\section*{Acknowledgments}

We thank T. Apollaro for discussions.
This work was supported by CNPq,
CNPq-Rede Nanobioestruturas,
CAPES, FINEP (Federal Brazilian Agencies), and 
FAPEAL (Alagoas State Agency). 

\appendix*

\section{\label{appendix} One-leg occupation probability}

Here we show how to obtain
the overall occupation probability over time
in one of the legs of the physical ladder chain [Eq. (\ref{overallP})] given $\delta_{n} = \delta$ and $\gamma_{n} = \gamma$ for all $n$ [cf. Eqs. (\ref{mainH1}) and (\ref{mainH2})].

Let us denote 
\begin{equation}
\ket{\lambda_{k,j}} = \sum_{n}v_{k,n}^{(j)}\ket{\mathbf{n}}^{(j)},
\end{equation}
such that it satisfies the eigenvalue equation  $H^{(j)}\ket{\lambda_{k,j}}=\lambda_{k,j} \ket{\lambda_{k,j}}$, with $k = 1,\ldots,N$. Now, 
given $J_{n,1} = J_{n,2}$, both sets of eigenstates ($j=1,2$)
feature the same spatial profile. Also note that $\lambda_{k,2} = \lambda_{k,1}+\delta$. 
Applying the interaction Hamiltonian [Eq. (\ref{mainH2})] thus yields $H_{I}\ket{\lambda_{k,1(2)}} = \gamma\ket{\lambda_{k,2(1)}}$. Thereby we end up with 
a series of independent dimer-like interactions between the normal modes of each leg and the total Hamiltonian of the system may be rewritten as $H=\sum_{k}H_{k}$, with
\begin{align} \label{dimerH}
H_{k} &= \lambda_{k,1}\ket{\lambda_{k,1}}\bra{\lambda_{k,1}} + \lambda_{k,2}\ket{\lambda_{k,2}}\bra{\lambda_{k,2}} \nonumber \\
&\,\,\,\,\,\,+ \gamma(\ket{\lambda_{k,1}}\bra{\lambda_{k,2}} + \mathrm{H.c.}).
\end{align}

Each dimer can be diagonalized separately and we get
\begin{equation}
\ket{\psi_{k}^{\pm}} = A^{\pm} \ket{\lambda_{k,1}} + B^{\pm}\ket{\lambda_{k,2}},
\end{equation}
with 
\begin{equation}
A^{\pm} = \frac{2\gamma}{\sqrt{(\delta \pm \Omega)^2+4\gamma^{2}}}, \,\,\,\, B^{\pm} = \frac{\delta \pm \Omega}{\sqrt{(\delta \pm \Omega)^2+4\gamma^{2}}},
\end{equation}
and corresponding eigenenergies 
\begin{equation}
E_{k}^{\pm} = \frac{1}{2}(\lambda_{k,1}+\lambda_{k,2} \pm \Omega) = \lambda_{k,1}+\frac{1}{2}(\delta \pm \Omega),
\end{equation}
where $\Omega = \sqrt{\delta^{2}+4\gamma^{2}}$ is the effective Rabi frequency.

Now, if we initialize the system in a linear combination of the form
$\ket{\psi(0)} = \sum_{k}a_{k}(0)\ket{\lambda_{k,1}}$,
the time-evolved state reads
\begin{align}
\ket{\psi(t)} &= U(t)\ket{\psi(0)} = \sum_{k, \nu = \pm}e^{-iE_{k}^{\nu} t}\ket{\psi_{k}^{\nu}}\bra{\psi_{k}^{\nu}} \psi(0) \rangle \nonumber \\
& = \sum_{\nu=\pm}(A^{\nu})^{2}e^{-i\left( \frac{\delta+\nu \Omega}{2}\right)t} \sum_{k}a_{k}(t)\ket{\lambda_{k,1}} \nonumber \\
 &\,\,\,\,\,\, + \sum_{\nu=\pm}A^{\nu}B^{\nu}e^{-i\left( \frac{\delta+\nu \Omega}{2}\right)t} \sum_{k}a_{k}(t)\ket{\lambda_{k,2}},
\end{align} 
where $a_{k}(t) = a_{k}(0)e^{-i \lambda_{k}^{(1)} t}$. Therefore, the wavefunction
evolves in time following the intrachain eigenspectrum -- which, recall, is the same for both legs --
with coefficients $a_{k}(t)$  
modulated by the sums in $\nu$ (see equation above). The overall occupation probability $P^{(1)}(t) \equiv \sum_{k}|\bra{\lambda_{k,1}} \psi(t)\rangle|^{2}$ can then be worked out as
\begin{equation}\label{overallP}
P^{(1)}(t) = 1- 2\left( \frac{\gamma}{\Omega} \right)^{2} \left[ 1-\cos(\Omega t) \right],
\end{equation}
which reduces to $P^{(1)}(t) = \cos^{2}(\gamma t)$ when $\delta = 0$. Likewise, $P^{(2)} = 1-P^{(1)}$
for the other leg.


%

\end{document}